\documentclass[acmtocl]{acmtrans2m}

\newtheorem{theorem}{Theorem}[section]

\newdef{definition}[theorem]{Definition}
\newdef{remark}[theorem]{Remark}

\title{\textit{P} is a proper subset of \textit{NP}.}
\author{JERRALD MEEK}

\begin{abstract}  
The purpose of this article is to examine and limit the conditions in which the $P$ complexity class could be equivalent to the \textit{NP} complexity class. Proof is provided by demonstrating that as the number of clauses in a \textit{NP-complete} problem approaches infinity, the number of input sets processed per computation performed also approaches infinity when solved by a polynomial time solution. It is then possible to determine that the only deterministic optimization of a \textit{NP-complete} problem that could prove \textit{P = NP} would be one that examines no more than a polynomial number of input sets for a given problem.\par
It is then shown that subdividing the set of all possible input sets into a representative polynomial search partition is a problem in the \textit{FEXP} complexity class. The findings of this article are combined with the findings of other articles in this series of 4 articles.  The final conclusion will be demonstrated that \textit{P} $\neq$ \textit{NP}.
\end{abstract}


\begin{document}

\begin{bottomstuff}
Several people who wish to remain anonymous have offered comments and suggestions which have improved this work. The author wishes to express his appreciation for their assistance.\par
\vspace{1ex}
The author would also like to thank Stephen Cook who identified an incorrect premise in a previous version of this article which was related to the nature of non determinism. This contribution resulted in a major revision of the article. Although Cook has granted permission to mention his contribution, this does not imply any endorsement.\par
	\begin{flushright}
		Jerrald Meek Copyright \copyright 2008
	\end{flushright}
\end{bottomstuff}

\maketitle

\section{Introduction.}
Stephen Cook described the inportance of the \textit{P = NP} question in his article \textit{The P Versus NP Problem} \protect\cite{cook_2}. Cook noted that if $P$ were to be proven equal to \textit{NP} then the consequences would be devastating to cryptography, yet Cook added ``it would also have stunning practical consequences of a more positive nature.'' These consequences could transform not only computer science, but mathematics in general.\par

Even if it turns out that $P\neq$ \textit{NP}, Cook hoped that ``every \textit{NP} problem [may be] susceptible to a polynomial-time algorithm that works on ``most'' inputs.'' \protect\cite[p. 6]{cook_2}\par

In this article it will be shown that as the number of clauses in a \textit{NP-complete} problem approaches infinity, the number of input sets processed per computation performed also approaches infinity when solved by a polynomial time solution. This will be used as the basis for proving the \textit{P = NP Optimization Theorem}, Theorem 4.4, which will be used to develop the \textit{P = NP Partition Theorem}, Theorem 5.1.\par

By the end of this article, the requirements for $P$ = \textit{NP} will be narrowed.  The narrowing of the requirements for $P$ = \textit{NP} will form the basis of following articles in this series, including:

\begin{enumerate}
	\addtocounter{enumi}{1}
	\item \textit{Analysis of the deterministic polynomial time solvability of the 0-1-Knapsack problem}. \protect\cite{meek2}
	\item \textit{Independence of P vs. NP in regards to oracle relativizations}. \protect\cite{meek3}
	\item \textit{Analysis of the postulates produced by Karp's Theorem}. \protect\cite{meek4}
\end{enumerate}

\vspace{2ex}
The results of these combined works will conclude that $P$ = \textit{NP} is unattainable.

\section{Preliminaries.}
The definition of a Deterministic Turing Machine will be based on that used by Marion \protect\cite[p. 61]{marion}.

\begin{definition}
\textbf{\emph{Deterministic Turing Machine.}}\par

A Deterministic Turing Machine $M$ has a finite set of \textit{states} $K$ with the \textit{state} $s$ being the \textit{initial state}. There exists a set $F$ with zero or more elements representing the \textit{final} or \textit{accepting states}, $F \subset K$. There is an alphabet $\Sigma$ which contains a symbol $B$ representing a blank. There exists an input alphabet $\Gamma$ such that $\Gamma \subset M-\left\{B\right\}$. There is a \textit{transition function} $\delta$ which defines an action to take depending on the current state and input. The Turing Machine
\[M=\left(K, \Sigma, \Gamma, \delta, s, F\right)\]
can recognize a language if $F\neq \oslash$, and $w$ is an input string (a finite sequence of elements from $\Gamma$). If the computation generated by the input string $w$ causes the machine to halt in a state that is an element of $F$ then $w$ is said to be accepted by $M$.
$L(M)$ consists of all input strings accepted by the machine.
\end{definition}

\begin{definition}
\textbf{\emph{Non-Deterministic Turing Machine.}}\par

The formula used by Marion to describe a Deterministic Turing Machine is the same formula used to describe a Non-Deterministic Turing Machine. Marion describes the difference as:

\begin{quotation}
\noindent Non-determinism lends an element of ``guessing'' to the process at each state where there is more than one choice the next transition is guessed. \protect\cite[p. 63]{marion}
\end{quotation}

Karp uses a definition of Non-Determinism that is similar to the concept of multithreading, where the machine performs a fork process to duplicate itself into multiple new machines with identical data, each executing a thread with a different option.

\begin{quotation}
\noindent A nondeterministic algorithm can be regarded as a process which, when confronted with a choice between (say) two alternatives, can create two copies of itself, and follow up the consequences of both courses of action. Repeated splitting may lead to an exponentially growing number of copies; the input is accepted if any sequence of choices leads to acceptance. \protect\cite[p. 91]{karp}
\end{quotation}

The definition from the \textit{National Institute of Science and Technology} \protect\cite{nist} is:

\begin{quotation}
\noindent A Turing machine which has more than one next state for some combinations of contents of the current cell and current state. An input is accepted if any move sequence leads to acceptance.
\end{quotation}

The most important thing to keep in mind about non determinism is that all possible input sets can be evaluated at the same time, but the machine only tells us if one input set evaluates \textit{true}. We therefore cannot expect a non deterministic machine to give us all inputs that evaluate \textit{true} in a single operation.
\end{definition}

\subsection{Why are \textit{NP-complete} problems so hard?}
The nature of a Boolean Satisfiability Problem (SAT) is to determine if an input exists that will result in the problem evaluating \textit{true}.\par

\noindent \begin{itemize}
	\item Let $A=\left\{ {a, b, c} \right\}$
	\item Let $B=\left\{ {d, e, f} \right\}$
	\item Let $C=\left\{ {g, h, i} \right\}$
	\item $x \in A$
	\item $y \in B$
	\item $z \in C$
	\item A 1-SAT problem is $x \wedge y \wedge z$
\end{itemize}

A logical conjunction can evaluate \textit{true} if and only if all literals are \textit{true}. It would therefore be easy to determine the truth of the expression if the values of $x$, $y$, and $z$ were given as \textit{true} or \textit{false}, but because they are each one of three options it is not as easy. With 3 options for each of 3 literals there are 27 possible inputs.

\[\left[a \wedge d \wedge g\right] \vee \left[b \wedge d \wedge g\right] \vee \left[c \wedge d \wedge g\right] \vee \left[a \wedge e \wedge g\right] \vee \left[b \wedge e \wedge g\right]\ldots \]

It happens to be easier to write the logically equivalent 3-SAT problem.

\[\left[a \vee b \vee c\right] \wedge \left[d \vee e \vee f\right] \wedge \left[g \vee h \vee i\right]\]

However, if we want to determine the values for $x$, $y$, and $z$ that will make the expression evaluate \textit{true}, it is still necessary to compare every possible value of $x$ to every possible value of $y$ and each of those combinations to every possible value of $z$.\par

If we are only looking to find one input set that evaluates \textit{true}, then optimization may be possible. This process could be optimized by first finding a value of $x$ that evaluates \textit{true}, then finding a value for $y$ that evaluates \textit{true}, and then finding a value for $z$ that evaluates \textit{true}. In this case the problem is solved in $3\times 3=9$ computations.\par

If we are looking to find all input sets that evaluate \textit{true}, then each value of $x$ that is \textit{true} must be found. Any values of $x$ that work must be duplicated for each \textit{true} $y$ value, and all of these sets must be duplicated for each \textit{true} $z$ value. If it turns out that several values of $x$, $y$, and $z$ are \textit{true} then there may not be any known polynomial time optimization for this problem. It is important to remember that a problem is proven irreducible to the $P$ complexity when it is impossible to find \textbf{one} answer in polynomial time on a deterministic machine.\par

Even the optimization for finding a single value can get complicated, especially when the literals have dependencies on each other.

\noindent \begin{itemize}
	\item Let $A = \left\{ {a, b, c} \right\}$
	\item Let $B = \left\{ {d, e, f} \right\}$
	\item Let $C = \left\{ {h, i, j} \right\}$
	\item Let $D = \left\{ {k, l, m} \right\}$
	\item Let $1 \leq n \leq 3$
	\item Let $1 \leq o \leq 3, o \neq n$
	\item Let $1 \leq p \leq 3, p \neq o, p \neq n$
	\item $w = A_n$
	\item $x = B_o$
	\item $y = C_p$
	\item $z \in D$
\end{itemize}

The problem will be
\[ w \wedge x \wedge y \wedge z \]

Now we have
\[\left[ {a \wedge e \wedge j \wedge k} \right] \vee \left[ {a \wedge e \wedge j \wedge l} \right] \vee \left[ {a \wedge e \wedge j \wedge m} \right] \vee \left[ {a \wedge f \wedge i \wedge k} \right] \vee\]
\[\left[ {a \wedge f \wedge i \wedge l} \right] \vee \left[ {a \wedge f \wedge i \wedge m} \right] \vee \left[ {b \wedge d \wedge j \wedge k} \right] \vee \left[ {b \wedge d \wedge j \wedge l} \right] \vee \ldots\]

If the values of $a$ and $e$ are both \textit{true}, and the algorithm arbitrarily decides to set $x = a$ and $y = e$. Then no accepted input may be found if $j$ is \textit{false}. This means the algorithm will have to start over again, assigning either $x$ or $y$ to a different value.\par

Therefore, it may be possible to optimize some \textit{NP-complete} problems with a deterministic algorithm specifically suited to work for a problem with a specific form. However, this optimization may not apply to all \textit{NP-complete} problems.

\subsection{How a Non-Deterministic Turing Machine evaluates\newline
\noindent \textit{NP-complete} problems.}
By Marion's definition, a Non-Deterministic Turing Machine will be lucky enough to guess the input that will cause the expression to evaluate \textit{true} on the first try if at least one such input exists.\par

By Karp's definition, a Non-Deterministic Turing Machine will branch into as many different Turing Machines as needed. These machines are coordinated to simultaneously evaluate one of each of the possibilities and find all inputs that cause the expression to evaluate \textit{true} if at least one exists. There should be no limit on the number of times that a Non-Deterministic Machine can branch.

\section{The input sets of \textit{NP-complete} problems.}
NP-complete is defined by Karp as:\par

\noindent The language $L$ is $(\underline{polynomial})$ $\underline{complete}$ if
\begin{tabbing}
	12345678\=\kill
	\> a) $L\in NP$ \\
	and \> b) SATISFIABILITY $\propto L$ \\
\end{tabbing}
\begin{quotation}
\noindent Either all complete languages are in $P$, or none of them are. The former alternative holds if and only if \textit{P = NP}. (Theorem 3 From: Reducibility among combinatorial problems) \protect\cite[p. 93]{karp}
\end{quotation}

\begin{definition}
\textbf{\emph{K-SAT}}\par

K-SAT is the Boolean Satisfiability Problem consisting of a conjunction of clauses, each being a disjunction of literals. K-SAT is \textit{NP-complete} when $k \geq 3$.
\end{definition}

\subsection{Total number of possible K-SAT input sets.}
\begin{itemize}
	\item Let a K-SAT problem have $k$ literals per clause.
	\item Let a K-SAT problem have $n$ clauses.
	\item Let $x$ be a two dimensional set containing $n$ sets each of which have $k$ elements.
	\item A CNF-K-SAT problem has the form:
		\[\left[ {x_{1_1} \vee x_{1_2} \vee x_{1_3} \vee \ldots \vee x_{1_k}} \right] \wedge\]
		\[\left[ {x_{2_1} \vee x_{2_2} \vee x_{2_3} \vee \ldots \vee x_{2_k}} \right] \wedge\]
		\[\left[ {x_{3_1} \vee x_{3_2} \vee x_{3_3} \vee \ldots \vee x_{3_k}} \right] \wedge\]
		\[ \vdots \]
		\[\left[ {x_{n_1} \vee x_{n_2} \vee x_{n_3} \vee \ldots \vee x_{n_k}} \right] \]
\end{itemize}

\begin{proof}
Clause 1 has $k$ literals. If there is a second clause then there are $k$ more literals, so there are $2k$ literals. If there is a third clause then there are $k$ more literals, totaling $3k$ literals. So the total number of literals is $nk$.\par

We could then say that $x$ is a one dimensional set with $kn$ elements. Each element can be \textit{true} or \textit{false}. We could then think of $x$ as a binary number with $kn$ digits.\par

It then follows that the number of posible input sets is $2^{kn}$.
\end{proof}

\subsection{Polynomial time computation rate of \textit{NP-complete} problems.}
\begin{itemize}
	\item Let $k$ be the number of literals in a clause such that $k\geq 3$.
	\item Let $n$ be the number of clauses in a \textit{NP-complete} class problem.
	\item Let $t(x)$ be a polynomial function representing the number of computations required for a problem in the complexity class \textit{NP-complete} to be solved in polynomial time.
	\item The number of input sets for a \textit{NP-complete} problem as shown in Section 3.1 is $2^{kn}$.
\end{itemize}

\begin{proof}
$r$ shall represent the number of input sets evaluated per computation performed.
	\[ r(n)=\frac{2^{kn}}{t(n)} \]

If a \textit{NP-complete} problem is solved in polynomial time by a search algorithm, then $r(n)$ represents that the polynomial time solution must evaluate $2^{kn}$ input sets for every $t(n)$ computations. This is the case provided that the method of solving the \textit{NP-complete} problem checks all possible input sets.
\end{proof}

\section{The limit of \textit{NP-complete} polynomial time computation rates.}
This section will rely on the following definitions.

\begin{definition}
\textbf{\emph{L'H\^{o}pital's Rule.}}\par

Let $f$ and $g$ be functions that are differentiable on an open interval $(a, b)$ 
containing $c$, except possibly at $c$ itself. Assume that ${g}'(x)\ne 0$ for all 
$x$ in $(a, b)$, except possibly at $c$ itself. If the limit of $f(x)/g(x)$ as $x$ approaches $c$ produces the indeterminate form 0/0, then
	\begin{tabbing}
		12345\=\+\kill
		$\mathop {\lim }\limits_{x\to c} \frac{f(x)}{g(x)}=\mathop {\lim }\limits_{x\to c} \frac{{f}'(x)}{{g}'(x)}$ \\
	\end{tabbing}
provided that the limit on the right exists (or is infinite). This result also 
applies if the limit of $f(x)/g(x)$ as $x$ approaches $c$ produces any one of the 
indeterminate forms $\infty /\infty $, $(- \infty )/\infty $, $\infty 
/(- \infty )$, or $(- \infty )/(- \infty )$. \protect\cite[p. 524]{calc}\par
\end{definition}

\begin{definition}
\textbf{\emph{An Infinite Limit at Infinity.}}\par

Let $f(x)$ be a function defined on an interval that contains $x=c$, except possibly at $x=c$. The statement
\begin{tabbing}
	12345\=\+\kill
	$\mathop {\lim }\limits_{x\to \infty } f(x)=\infty$ \\
\end{tabbing}
means that for each $N>0$ there exists a $M>0$ such that
\begin{tabbing}
	12345\=\+\kill
	$f(x)>N$ whenever $x>M$ \\
\end{tabbing}
\end{definition}

\begin{definition}
\textbf{\emph{Polynomial Time.}}\par
Cook defines polynomial time as, ``We say that $M$ runs in \textit{polynomial time} if there exists $k$ such that for all $n$, $T_{M}(n)\leq n^{k} + k$.'' \protect\cite[p. 1]{cook_2}\par

Cook's definition will be used here because it will allow us to assume that polynomial time is always bounded by a function that is the sum of 2 monomials. It would be perfectly acceptable for the time to be bounded by the sum of 3 or more monomials, but that would only serve to complicate things. If it is accepted that $\left(n^k + k\right) > \left(n^{k-1} + n^{k-2} + k\right)$, then Cook's definition will always work.\par

In this article Cook's definition will be altered to the following: there exists $p$ such that for all $n$, $T_{M}(n)\leq an^p + p$, when $a > 0$. This is because $k$ is already being used to represent the number of literals per clause. While $p$ will probably be proportional to $k$, the statement $p = k$ can only be known if the algorithm is known. In this work, nothing is being said about any polynomial time algorithm other than that it is executed in no more than $an^{p} + p$ computations, and it is a function that solves a \textit{NP-complete} problem.
\end{definition}

\subsection{Exponential functions $>$ polynomial functions.}
	\begin{itemize}
		\item Let $a$ be a constant such that $a > 0$.
		\item Let $p$ be a constant such that $p > 0$.
		\item $c= \left( \ln \;a \right)$ which is also a constant.
		\item $f(x)=a^x$
		\item $g(x)=ax^{p} + p$
	\end{itemize}
\begin{proof}
A pattern will be demonstrated by taking the first 3 derivatives of $f(x)$ and $g(x)$:

	\begin{tabbing}
		12345\=\+\kill
		${f}'(x) = \frac{d}{dx} \left[ {a^x} \right] = \left( \ln \; a \right) a^x = ca^x $ \\
		${f}''(x) = \frac{d}{dx} \left[ {ca^x} \right] = c \left( \ln \; a \right) a^x = c^2a^x $ \\
		${f}'''(x) = \frac{d}{dx} \left[ {c^2a^x} \right] = c^2 \left( \ln \;a \right) a^x = c^3a^x $ \\
	\end{tabbing}
	\begin{tabbing}
		12345\=\+\kill
		${g}'(x)=\frac{d}{dx}\left[ {ax^p+p} \right]=pax^{p-1}+0=pax^{p-1} $ \\
		${g}''(x)=\frac{d}{dx}[pax^{p-1}]=p\left( {p-1} \right)ax^{p-2} $ \\
		${g}'''(x)=\frac{d}{dx}\left[ {p\left( {p-1} \right)ax^{p-2}} 
\right]=p\left( {p-1} \right)\left( {p-2} \right)ax^{p-3} $ \\
	\end{tabbing}

As can be seen from this pattern, the $(p-1)^{th}$ derivative of $f(x)$ would be $c^{p-1}a^x$, while the $(p-1)^{th}$ derivative of $g(x)$ will be a constant.\par

L'H\^{o}pital's Rule can be expanded as follows.
	\begin{tabbing}
		12345\=\+\kill
		$\mathop {\lim }\limits_{x\to c} \frac{f(x)}{g(x)}=\mathop {\lim }\limits_{x\to c} \frac{{f}'(x)}{{g}'(x)}=\mathop {\lim }\limits_{x\to c} \frac{{f}''(x)}{{g}''(x)}=\mathop {\lim }\limits_{x\to c} \frac{{f}'''(x)}{{g}'''(x)}$ \\
	\end{tabbing}

Let $e(x)$ be the $(p-1)^{th}$ derivative of $f(x)$ and let $h(x)$ be the $(p-1)^{th}$ derivative of $g(x)$. Because it is known that the $(p-1)^{th}$ derivative of $g(x)$ will not be zero, we can say
	\begin{tabbing}
		12345\=\+\kill
		$\mathop {\lim }\limits_{x\to c} \frac{f(x)}{g(x)}=\mathop {\lim }\limits_{x\to c} \frac{e(x)}{h(x)}$ \\
	\end{tabbing}

That is,

	\begin{tabbing}
		12345\=\+\kill
		$\mathop {\lim }\limits_{x\to \infty} \frac{f(x)}{g(x)}=\mathop {\lim }\limits_{x\to \infty} \frac{c^{p-1}a^x}{CONSTANT}=\infty $ \\
		\\
		$\mathop {\lim }\limits_{x\to \infty} \frac{f(x)}{g(x)}=\infty \rightarrow {f(x)} > {g(x)}$
	\end{tabbing}
For any set of functions $\left\{ {f(x), g(x)} \right\}$ in which $f(x)$ is exponential and 
$g(x)$ is polynomial, there exists a number $l$ such that any number $n\ge l$ will 
make the statement \textit{true} that $f(n) > g(n)$.
\end{proof}

\subsection{Limit at infinity of polynomial time computation rates for \textit{NP-complete} problems.}
	\begin{itemize}
		\item Let $k$ be the number of literals in a clause such that $k\geq 3$.
		\item Let $n$ be the number of clauses in a NP-complete class problem.
		\item Let $t(n)$ be a polynomial function representing the number of computations
required for a problem in the complexity class \textit{NP-complete} to be solved in
polynomial time.
		\item If a \textit{NP-complete} class problem is solved in polynomial time, then the number of input sets processed by the polynomial function per computation performed as shown in Section 3.2 is $r(n)=\frac{2^{kn}}{t(n)}$.
	\end{itemize}
\begin{proof}
Let $a$ be a number at which $2^{ak}>t(a)$. Section 4.1 indicates such a 
number must exist.

\begin{table}[h]
\begin{tabular}{p{171pt}l}
$\mathop {\lim }\limits_{n\to \infty } \frac{2^{kn}}{t(n)}=\infty $& 
Assume the limit of $r(n)$ is at infinity. \vspace{3pt} \\
$\frac{2^{kn}}{t(n)}>N\leftarrow n>M$& 
Definition of an Infinite limit at Infinity. \vspace{3pt} \\
$\frac{2^{kn}}{t(n)}>\raise0.7ex\hbox{$1$} \!\mathord{\left/ {\vphantom {1 {t(n)}}}\right.\kern-\nulldelimiterspace}\!\lower0.7ex\hbox{${t(n)}$}\leftarrow n>a$& 
Set N equal to $\raise0.7ex\hbox{$1$} \!\mathord{\left/ {\vphantom {1 {t(n)}}}\right.\kern-\nulldelimiterspace}\!\lower0.7ex\hbox{${t(n)}$}$ and M equal to $a$. \vspace{3pt} \\
$\frac{t(n)2^{kn}}{t(n)}>\raise0.7ex\hbox{${t(n)}$} \!\mathord{\left/ {\vphantom {{t(n)} {t(n)}}}\right.\kern-\nulldelimiterspace}\!\lower0.7ex\hbox{${t(n)}$}\leftarrow n>a$& 
Multiply both sides by $t(n)$.  \vspace{3pt} \\
$2^{kn}>1\leftarrow n>a$& 
Cancel like terms. \\
\end{tabular}
\end{table}

\[\mathop {\lim }\limits_{n\to \infty } r(n)=\infty\]

Therefore, as the number of clauses in a \textit{NP-complete} problem increases, the number of input sets that must be processed per computation performed will eventually exceed any finite limit.
\end{proof}

\subsection{The limitation of \textit{NP-complete} optimizations.}
A Deterministic Turing Machine is limited to checking no more than one input per computation. Section 4.2 shows that when a polynomial time algorithm is used to check all possible input sets for a \textit{NP-Complete} problem, the machine cannot be limited by the number of inputs checked per computation.\par

\begin{theorem}
\textbf{\emph{P = NP Optimization Theorem.}}\par

The only deterministic optimization of a \textit{NP-complete} problem that could prove \textit{P = NP} would be one that can always solve a \textit{NP-complete} problem by examining no more than a polynomial number of input sets for that problem.
\end{theorem}

\section{The dificulty of creating a polynomial optimization.}
It may be possible to optimize a deterministic algorithm so that a solution to a \textit{NP-complete} problem may be found in polynomial time, but doing so would require that the optimization must limit the number of inputs checked to a polynomial number of sets. Therefore, we can assume that as the number of clauses in the problem approaches infinity, the percentage of possible inputs checked by the polynomial time algorithm would approach zero. This could be acceptable if\newline
\begin{enumerate}
	\item The nature of the problem is such that all input sets not checked are guaranteed to cause the expression to evaluate as \textit{false}.\newline
		\begin{center}
		or
		\end{center}
	\item The nature of the problem is such that if one or more input sets that cause the expression to evaluate \textit{true} exists, then at least one of those sets will exist in the set that is checked.
\end{enumerate}

Any optimization technique that relies on reducing the number of input sets checked to a polynomial amount can be expected not to work for all \textit{NP-complete} problems. This can be demonstrated as follows:\par
\noindent \begin{itemize}
	\item Let $A$ be a set containing all possible input sets for \textit{NP-complete} problems $f(x)$ and $g(x)$.
	\item $B \subset A$
	\item $C = A - B$
	\item Let $f(x)$ be an \textit{NP-complete} problem that evaluates \textit{false} when the input is an element of $B$.
	\item Let $g(x)$ be an \textit{NP-complete} problem that evaluates \textit{false} when the input is an element of $C$.
\end{itemize}

Suppose that an optimized algorithm for $f(x)$ is found that will not evaluate elements of $B$. This technique will find the solution faster because it is already known that checking elements of $B$ for $f(x)$ is a waste of time. However this algorithm will always fail for evaluating $g(x)$.\par

It might be possible to create a different optimization that will evaluate a subset of $B$ and a subset of $C$ in such a way that it will always find at least one accepting input set for both $f(x)$ and $g(x)$ if one exists. In this case it should be possible to define a new problem $e(x)$ that is \textit{NP-complete} and has at least one input set causing the expression to evaluate \textit{true}, but none in the range checked by the optimized algorithm for solving $f(x)$ or $g(x)$.\par

\subsection{Polynomial time under \textit{P = NP} limitations.}
In this section the process of solving a \textit{NP-complete} problem by only examining a polynomial-sized partition of the set of all possible solutions will be examined. The term ``representative polynomial search partition'' will be used to indicate a partition from the set of all possible input sets such that the partition has a polynomial cardinality, and will contain at least one input set that results in a \textit{true} evaluation if such an input set exists.

\subsubsection{The Form of a \textit{P = NP} algorithm}
The \textit{P = NP Optimization Theorem} requires that a deterministic polynomial time algorithm for a \textit{NP-complete} problem must perform a search limited to a polynomial-sized subset of all possible input sets. It is also required that at least one input set that results in a \textit{true} evaluation, if such a set exists, must exist within the representative polynomial search partition. It should then be a reasonable assumption that if a partition meeting these criteria must be searched, then it must first be found.\par

\noindent \begin{itemize}
	\item Let $C$ be a \textit{NP-complete} problem.
	\item Let $S$ be the set of all possible input sets for $C$.
	\item Let $A \subset S$ such that $A$ contains all elements of $S$ that results in a \textit{true} evaluation for $C$.
	\item Let $P \subset S$ such that $P$ contains a polynomial number of elements.
	\item Let $\alpha$ be a set such that $\alpha = \oslash \Leftrightarrow A = \oslash$ and $P \supset \alpha \subset A$
	\item Let $t_e$ be the number of computations required for a single element of $P$ to be evaluated.
	\item Let $t_p$ be the number of computations required for a single element of $P$ to be found.
	\item Let $t_{max}$ be the maximum number of computations required to solve $C$.
\end{itemize}

\begin{proof}
The \textit{NP-Complete Optimization Theorem} requires that any deterministic polynomial time algorithm for $C$ must have the form
\noindent \begin{enumerate}
	\item Set $i = 1$
	\item Find $P_i$
	\item If $P_i \in \alpha$ then halt in an accepting state.
	\item If $P_i \notin \alpha$ and $i < \left| {P} \right|$ then increment $i$ and continue at step 2.
	\item If $P_i \notin \alpha$ and $i = \left| {P} \right|$ then halt in a non-accepting state.
\end{enumerate}

This algorithm requires the longest time when $A = \oslash$, or when $\left| {\alpha} \right| = 1$ and $P_{\left| {P} \right|} \in \alpha$. In both cases the algorithm will be forced to find all elements of $P$. It is then the case that

\[t_{max} = \left| {P} \right| \left( {t_p + t_e} \right) \]

It is known that $\left| {P} \right|$ and $t_e$ are products of polynomial functions. If \textit{P = NP} then $t_{max}$ must also be a product of a polynomial function. It is then the case that $t_p$ must be a polynomial function.
\end{proof}

\subsubsection{The complexity of finding the representative polynomial search partition}
\noindent \begin{itemize}
	\item Let $C$ be a \textit{NP-complete} problem.
	\item Let $S$ be the set of all possible input sets for $C$.
	\item Let $A \subset S$ such that $A$ contains all elements of $S$ that result in a \textit{true} evaluation for $C$.
	\item Let $q$ be some quality possessed by only a polynomial number of elements from $S$.
	\item Let $P \subset S$ such that $P$ contains all elements of $S$ which possess quality $q$.
	\item Let $\alpha$ be a set such that $\alpha = \oslash \Leftrightarrow A = \oslash$ and $P \supset \alpha \subset A$
\end{itemize}
\begin{proof}
It should be a reasonable assumption that all elements of $P$ share some quality in common, which is absent in all elements of $S$ that are not elements of $P$. If this were not the case, then it would be impossible to discriminate the polynomial number of elements of $P$ from the exponentially many other elements of $S$.

The algorithm for finding all elements of $P$ by exhaustion is
\noindent \begin{enumerate}
	\item Set $i = 1$
	\item If $q \mapsto S_i$ then $S_i \in P$
	\item If $i < \left| {S} \right|$ then increment $i$ and continue at step 2.
	\item If $i = \left| {S} \right|$ then all elements of $P$ have been found.
\end{enumerate}

Notice that the algorithm iterates once for every element of $S$. It is then the case that the process of finding all elements of $P$ requires an exponential number of iterations. It should be expected that an exponential number of iterations will require an exponential number of computations.\par

This problem is then a member of a complexity class that could be described as \textit{FEXP}. This is a function problem that requires exponential time on a Deterministic Turing Machine.
\end{proof}

\subsubsection{$P \neq$ \textit{NP} when the representative polynomial search partition is found by exhaustion}
\noindent \begin{itemize}
	\item Let $C$ be a \textit{NP-complete} problem.
	\item Let $t_s$ be the maximum number of computations required to solve $C$ by the exhaustion method.
	\item Let $t_p$ be the number of computations required to find a representative polynomial search partition for $C$ by the exhaustion method.
\end{itemize}

\begin{proof}
If a \textit{NP-complete} problem is solved in deterministic exponential time, then the algorithm may iterate through every element of the set of all possible inputs. However, the algorithm also can stop searching when the algorithm finds an input set that evaluates \textit{true}. Therefore, a deterministic exponential time algorithm for a \textit{NP-complete} problem may or may not require the entire duration of its worst case run time.\par

In the previous section it was shown that the process for finding the elements of the representative polynomial search partition by exhaustion requires searching all elements of the set of all possible inputs. Finding all elements of the search partition does not end when an element is found, because there may be more. It is therefore the case that the algorithm for finding all elements of the representative polynomial search partition must always require the entire duration of its worst case time (if the entire partition is found).\par

It may also be of interest to mention that the process of recording those elements of the representative polynomial search partition that are found will inevitably require even more computations. However, even if this is ignored it still remains that
\[ t_p \geq t_s \]
\end{proof}

\subsection{The limitation of \textit{NP-complete} search partitioning.}
Section 5.1.1 shows that a deterministic polynomial time solution for a \textit{NP-complete} problem must produce a representative polynomial search partition in deterministic polynomial time. Section 5.1.3 indicates that evaluating all elements of the set of all possible inputs to find elements of the search partition results in a \textit{NP-complete} problem being solved in deterministic exponential time.

\begin{theorem}
\textbf{\emph{P = NP Search Partition Theorem.}}
The only deterministic search optimization of a \textit{NP-complete} problem that could prove \textit{P = NP} would be one that can always find a representative polynomial search partition by examining no more than a polynomial number of input sets from the set of all possible input sets.
\end{theorem}

\section{Examples of \textit{NP-complete} solutions.}
The set of all possible input sets for a \textit{NP-complete} problem is exponential in size. If a deterministic polynomial time algorithm is to be found, then a representative polynomial search partition must be found in deterministic polynomial time. For a search partition to be found in deterministic polynomial time, then a search partition for that search partition must be found in deterministic polynomial time. This circular argument means that finding a deterministic polynomial time algorithm for a \textit{NP-complete} problem can be done only if a deterministic polynomial time algorithm for that problem already exists.

\subsubsection{A \textit{NP-complete} problem solved in deterministic polynomial time}
Any problem solvable in deterministic polynomial time can also be solved in polynomial time on a Non Deterministic Turing Machine. It is therefore the case that any problem in the complexity class of $P$ is also a member of the complexity class of \textit{NP}. All problems in \textit{NP} can be reduced to \textit{NP-complete}, although some problems in \textit{NP-complete} appear not to be reducible to problems outside \textit{NP-complete}. Because all problems in $P$ are also in \textit{NP}, and all problems in \textit{NP} are reducible to \textit{NP-complete}, it then follows that all problems in $P$ are reducible to \textit{NP-complete}.\par

In this section a $P$ class problem will be represented as a \textit{NP-complete} problem. The purpose of this demonstration is to examine how the trap of the \textit{P = NP Search Partition Theorem} may be avoided.

\noindent \begin{itemize}
	\item Let $S$ be a set of real numbers with $k$ elements such that $k \geq 3$
	\item Let $M$ be a set of real numbers with $n$ elements such that $n \leq k$
\end{itemize}

\begin{proof}
The \textit{NP-complete} problem represents the question, is $M$ a subset of $S$? This problem is formulated by:
\[ \left[ {S_1 = M_1 \vee S_2 = M_1 \vee S_3 = M_1 \vee \ldots \vee S_k = M_1} \right] \wedge \]
\[ \left[ {S_1 = M_2 \vee S_2 = M_2 \vee S_3 = M_2 \vee \ldots \vee S_k = M_2} \right] \wedge \]
\[ \left[ {S_1 = M_3 \vee S_2 = M_3 \vee S_3 = M_3 \vee \ldots \vee S_k = M_3} \right] \wedge \]
\[ \vdots \]
\[ \left[ {S_1 = M_n \vee S_2 = M_n \vee S_3 = M_n \vee \ldots \vee S_k = M_n} \right] \]

The set of all possible input sets for this problem has $2^{kn}$ elements. However, if both $S$ and $M$ are sorted in ascending order, then $M_1$ can be compared to every value of $S$ in ascending order. If a match is found for $M_1$, then $M_2$, which is greater than $M_1$, need not be compared to any values of $S$ less than $M_1$. This method could be used to determine if $M$ is a subset of $S$ in $k$ computations.\par

Notice that this method will find a match if one exists, and it will not if one does not exist. It could be said that this algorithm finds a search partition with 1 element in deterministic polynomial time if $M$ is a subset of $S$, or it fails to find a search partition if $M$ is not a subset of $S$.
\end{proof}

\subsubsection{A \textit{NP-complete} problem unsolvable in deterministic polynomial time}
The previous example allows a problem to be solved in polynomial time because the nature of the problem allows a search partition to be found in polynomial time. Horowitz and Sahni \protect\cite{horowitz} have applied a similar method to the \textit{Knapsack} problem. The form of the \textit{Knapsack} problem examined here is the \textit{0-1-Knapsack} problem.\par
\noindent \begin{itemize}
	\item Let $S$ be a set of real numbers with no two identical elements.
	\item Let $r$ be the number of elements in $S$.
	\item Let $\delta$ be a set with $r$ elements such that
		\[ {\delta}_i \in \left\{ {0, 1} \right\} \leftarrow 1 \leq i \leq r \]
	\item Let $M$ be a real number.
\end{itemize}

Then

\[ \sum_{i=1}^r {\delta}_i S_i = M \]

If given the values for $S$ and $M$, is it possible to create an algorithm guaranteed to find at least one variation of $\delta$ such that the expression will evaluate \textit{true} if any such variation exists, while only examining a polynomial subset of the total set of all possible variations for $\delta$?\par

\begin{proof}
This problem could be optimized by examining subsets of $S$. When $\sigma$ is a subset of $S$ found to be greater than $M$, then any subset known to be greater than $\sigma$ need not be examined. Horowitz and Sahni \protect\cite{horowitz} have found that this method still requires exponential time. This is because the method of Horowitz and Sahni cannot reduce the search partition to a polynomial-size.\newline
\noindent \begin{itemize}
	\item Let $a < b < c < d$
	\item Let $S = \left\{ {a, b, c, d} \right\}$
\end{itemize}

The sum of each subset of $S$ is

\begin{acmtable}{343pt}[h]
	\centering
	\begin{tabular}{|l|l|l|l|l|l|}
		1) 0 & 2) $a$ & 3) $b$ & 4) $c$ & 5) $d$ & \\
		\hline
		6) $a + b$ & 7) $a + c$ & 8) $a + d$ & 9) $b + c$ & 10) $b + d$ & 11) $c + d$ \\
		\hline
		12) $a + b + c$ & 13) $a + b + d$ & 14) $a + c + d$ & 15) $b + c + d$ & & \\
		\hline
		16) $a + b + c + d$ & & & & & \\
	\end{tabular}
	\caption{Sums of subsets of $S$.}
\end{acmtable}

If it is determined that all subsets that contain $d$ as an element cannot total to $M$, then the input sets 5, 8, 10, 11, 13, 14, 15, and 16 can be removed. This leaves $2^3 = 8$ input sets. So eliminating the exponential partition of all subsets containing any given element results in the problem still having an exponential number of input sets to be searched.\par

It may be possible to eliminate all elements that contain both $a$ and $b$. However, it is already known that the search partition resulting from removing all elements containing $a$ is exponential in size. The number of elements that contain both $a$ and $b$ is less than the number of elements that contain $a$. It should then be expected that this method will not result in a polynomial-sized search partition.\par

If input set 14 is found to be greater than $M$, then input sets 15 and 16 can also be eliminated because they will be greater than $M$. However, this does not reduce the maximum time required for the search. If $M = a + b + c + d$ and the algorithm searches in ascending order, it will still require the same amount of time.\par

There also have been attempts by Horowitz and Sahni \protect\cite{horowitz} to split the problem into several partitions. The partitioning methods of Horowitz and Sahni do provide faster methods, but they rely on exponential partitions and still require exponential time.\par

The problem with these kinds of reductions is that an exponential function minus a polynomial function equals an exponential function. If an exponential function is subtracted from an exponential function, then the result may be a polynomial function, or it may be an exponential function. Therefore, any relation that allows a polynomial-sized partition to be eliminated will not work, and any relation that allows an exponential-sized partition to be eliminated may or may not work. It happens to be that there is no easy way to find a representative polynomial search partition simply by examining the elements of $S$ and the value of $M$. A more detailed examination of the Knapsack problem in reguards to polynomial time solutions based upon the form of the problem can be found in \textit{Analysis of the Deterministic Polynomial Time Solvability of the 0-1-Knapsack Problem.} \protect\cite{meek2} and also in \textit{Analysis of the postulates produced by Karp's Theorem.} \protect\cite{meek4}\par

If the nature of the problem does not allow for a simple reduction to a representative polynomial search partition in deterministic polynomial time, then the conundrum of the \textit{P = NP Search Partition Theorem} prevents the problem from being solved in deterministic polynomial time.
\end{proof}

\section{Conclusion.}
The final conclusion is drawn from the following premises:
\begin{enumerate}
	\item The complexity class of $P$ is the class of problems that can be solved in polynomial time by a Deterministic Turing Machine.
	\item The \textit{P = NP Optimization Theorem} requires that an algorithm that solves a \textit{NP-complete} problem in deterministic polynomial time can examine no more than a polynomial-sized search partition, and must find this partition in deterministic polynomial time.\par
	\item Some \textit{NP-complete} problems can only have a deterministic polynomial time solution if the SAT problem has a deterministic polynomial time solution.  \protect\cite{meek2}
	\item SAT does not have a deterministic polynomial time solution.  \protect\cite{meek4}
\end{enumerate}

\subsection{\textit{P} is a proper subset of \textit{NP}.}
Because some \textit{NP-complete} problems are dependant upon SAT to produce a deterministic polynomial time solution for them, and because SAT does not have a deterministic polynomial time solution, then $P$ is a proper subset of \textit{NP}. $P \neq$ \textit{NP}.\par

\begin{flushright}
Q.E.D.
\end{flushright}

\section{Version history.}
The author has decided to include a record of the version history for this article. The motive of this history is to acknowledge the fact that after the author wrote the original draft of this article, it has taken on an organic life. The article has been revised several times. Although the central idea behind the article has changed only slightly, the means of explanation and methods of proof have transformed and evolved. Some changes have been the result of the author's own desire to clarify the argument, and other changes have been due to third party comments, suggestions and criticisms.\par

The author wishes to encourage further feedback which may improve, strengthen, or perhaps disprove the content of this article. For that reason the author does not publish the names of any specific people who may have suggested, commented, or criticized the article in such a way that resulted in a revision, unless premission has been granted to do so.\par

\noindent \textbf{arXiv Current Version}\newline
6Sep08 Submitted to arXiv.
\begin{itemize}
	\item Refrance to \protect\cite{meek3} added.
	\item Revision of introduction, conclusion, and abstract.
	\item Minor formatting changes.
\end{itemize}

\noindent \textbf{arXiv Version 11}\newline
23Aug08 Submitted to arXiv.
\begin{itemize}
	\item Refrance to \protect\cite{meek4} added.
\end{itemize}

\noindent \textbf{arXiv Version 10}\newline
09May08 Submitted to arXiv.
\begin{itemize}
	\item Minor corrections.
\end{itemize}

\noindent \textbf{arXiv Version 9}\newline
05May08 Submitted to arXiv.
\begin{itemize}
	\item Reformated for future submission to ACM.
	\item Refrance to \protect\cite{meek2} added.
\end{itemize}

\noindent \textbf{arXiv Version 8}\newline
24Apr08 Submitted to arXiv.
\begin{itemize}
	\item Minor corrections based off of suggestions from a proof reading.
	\item Two introductory paragraphs were added to \textit{A NP-Complete Problem Solved in Deterministic Polynomial Time}.
\end{itemize}

\noindent \textbf{arXiv Version 7}\newline
22Apr08 Submitted to arXiv. Major revision.
\begin{itemize}
	\item A more formal approach was adopted for proving that the \textit{P = NP Optimization Theorem} implies $P \neq$ \textit{NP}. The logic remains fundamentally the same; the new description is intended to provide an easier way of identifying any flaws in the argument.
	\item An example of a \textit{NP-complete} problem solvable in deterministic polynomial time was added.
\end{itemize}

\noindent \textbf{arXiv Version 6}\newline
18Apr08 Submitted to arXiv.
\begin{itemize}
	\item Minor corrections.
\end{itemize}

\noindent \textbf{arXiv Version 5}\newline
\noindent 16Apr08 Submitted to arXiv. This version represents a major change due to an invalid theorem in the previous version.
\begin{itemize}
	\item The proof of the \textit{P = NP Optimization Theorem} was confirmed by expert review.
	\item The proof of the \textit{NP-Complete Optimization Theorem} was disproved by expert review. As a result that theorem was abandoned.
	\item The \textit{P = NP Partitioning Time Theorem} was introduced and is awaiting confirmation or disproof.
\end{itemize}

\noindent \textbf{arXiv Version 4}\newline
\noindent 10Apr08 Submitted to arXiv.
\begin{itemize}
	\item \textit{The 1-SAT and 2-SAT Exemptions} were added after the author withdrew the article from publication during a panic induced by the mistaken belief that the argument required 1-SAT and 2-SAT to be \textit{NP-complete}. As a result of this incident, the author has chosen to delay any further submissions for publication.
\end{itemize}

\noindent \textbf{arXiv Version 3}\newline
\noindent 08Apr08 Submitted to arXiv.
\begin{itemize}
	\item The conclusion for \textit{Limit At Infinity of Polynomial Time Computation Rates for
NP-Complete Problems} was revised. This was the result of an email from an arXiv user.
	\item An extra paragraph was added to the \textit{Knapsack Can Fit Problem} for clarification.
\end{itemize}

\noindent \textbf{arXiv Version 2}\newline
\noindent 08Apr08 Submitted to arXiv.
\begin{itemize}
	\item An error was corrected in the \textit{Total number of possible K-SAT Input Sets} section. This was the result of an email from an arXiv user.
\end{itemize}

\noindent \textbf{arXiv Version 1}\newline
\noindent 07Apr08 Submitted to arXiv.
\begin{itemize}
	\item Explanation of the proof for the \textit{NP-Complete Optimization Theorem} was improved.
\end{itemize}

\noindent \textbf{Unpublished Version 0.3}\newline
\noindent 04Apr08 Submitted for publication to The \textit{New York Journal of Mathematics}, replacing previous submission. This submission was withdrawn by the author on 10Apr08 because the author mistakenly thought that the proof had a flaw that would require 1-SAT and 2-SAT to be \textit{NP-complete} for the argument to be valid. The author discovered upon closer examination that the theorem actually does allow for polynomial solutions to these problems.
\begin{itemize}
	\item The author found that the \textit{NP-Complete Optimization Conjecture} could be proven with the work of Horowitz and Sahni\protect\cite{horowitz}, and the conjecture was turned into a theorem.
	\item The conclusion was restated as it originally was.
\end{itemize}

\noindent \textbf{Unpublished Version 0.2}\newline
\noindent 03Apr08 Submitted for publication to The \textit{New York Journal of Mathematics}.
\begin{itemize}
	\item Author revised method of presentation. This was the result of an email from the AMS referee.
	\item The \textit{NP-Complete Optimization Conjecture} was introduced: the conclusion was altered to state \textit{P = NP} if and only if the \textit{NP-Complete Optimization Conjecture} is \textit{false}. This was the result of an email from the AMS referee.
\end{itemize}

\noindent \textbf{Unpublished Version 0.1}\newline
\noindent 28Mar08 Submitted for publication to \textit{AMS Journal Mathematics of Computation}. Rejected 1Apr08.

\begin{received}
Received xx/2008; revised xx/2008; accepted xx/2008
\end{received}
\end{document}